\documentclass[twocolumn,prd,floatfix,showpacs]{revtex4}
\usepackage{graphicx}
\usepackage{epsfig}

\begin{document}

\title{What is the lowest possible reheating temperature?}
\author{Steen Hannestad}
\affiliation{Department of Physics, University of Southern Denmark,
Campusvej 55, DK-5230 Odense M, Denmark \\
and
\\
NORDITA, Blegdamsvej 17, DK-2100 Copenhagen, Denmark}
\email{hannestad@fysik.sdu.dk}
\date{11 March 2004}

\begin{abstract}
We study models in which the universe exits reheating at
temperatures in the MeV regime. By combining light element
abundance measurements with cosmic microwave background and large
scale structure data we find a fairly robust lower limit on the
reheating temperature of $T_{\rm RH} \gtrsim 4$ MeV at 95\% C.L.
However, if the heavy particle whose decay reheats the universe
has a direct decay mode to neutrinos, there are some small islands
left in parameter space where a reheating temperature as low as 1
MeV is allowed. The derived lower bound on the reheating
temperature also leads to very stringent bounds on models with $n$
large extra dimensions. For $n=2$ the bound on the
compactification scale is $M \gtrsim 2000$ TeV, and for $n=3$ it
is 100 TeV. These are currently the strongest available bounds on
such models.

\end{abstract}

\maketitle

%%%%%%%%%%%%%%%%%%%%%%%%%%%%%%%%%%%%%%%%%%%%%%%%%%%%%%%%%%%%%%%%%%%%%%
%%%% Section I %%%%%%%%%%%%%%%%%%%%%%%%%%%%%%%%%%%%%%%%%%%%%%%%%%%%%%%
%%%%%%%%%%%%%%%%%%%%%%%%%%%%%%%%%%%%%%%%%%%%%%%%%%%%%%%%%%%%%%%%%%%%%%

\section{Introduction}

The standard big bang model has been tested thoroughly up to
temperatures around 1 MeV where big bang nucleosynthesis occurred.
At much higher temperatures the universe is assumed to have
undergone inflation, during which the primordial density
perturbations are produced.

Towards the end of inflation the inflaton potential steepens so
that slow roll is violated, and the universe enters the reheating
phase. During this phase all particles which are kinematically
allowed are produced, either by direct decay or from the thermal
bath produced by the inflaton decay.

Finally the universe enters the radiation dominated phase at a
temperature $T_{\rm RH}$, which is a function of the inflaton
decay rate. The only certain bound on this reheating temperature
comes from big bang nucleosynthesis, and has in several previous
studies been found to be around 1 MeV \cite{kolb,kks}.

It should be noted that even if the reheating temperature after
inflation is much higher there can still be subsequent "reheating"
phases, in the sense that reheating is defined to be a period
where the energy density is dominated by an unstable
non-relativistic particle species. In standard reheating this is
the inflaton, but in supersymmetric models it could for instance
be the gravitino.

In the present paper we update previous calculations of this
reheating phenomenon, using data from cosmic microwave background
and large scale structure observations. Furthermore we extend the
analysis to include the possibility of having a direct decay mode
of the heavy particle into light neutrinos. If the heavy particle
is a scalar this decay is normally suppressed by a factor
$(m_\nu/m_\phi)^2$ because of the necessary helicity flip.
However, the heavy particle could either be a non-scalar particle,
or it could be a pseudo-scalar like the majoron which couples only
to neutrinos. Even though such models are slightly contrived it is
of interest to study whether the temperature bound on reheating is
significantly affected by the possibility of direct decay into
neutrinos.

In section II we discuss the set of Boltzmann equations necessary
to follow the evolution of all particle species. In section III
present results of the numerical solution of these equation, and
in section IV we compare model predictions with observational
data. Finally, section V is a review of other astrophysical
constraints on heavy, decaying particles, and section VI contains
a discussion.

\section{Boltzmann equations}

We follow the evolution of all particles by solving the Boltzmann
equation for each species
\begin{equation}
\frac{\partial f}{\partial t} - H p \frac{\partial f}{\partial p}
= C_{\rm coll}, \label{eq:boltz}
\end{equation}
where $C_{\rm coll}$ is the collision operator describing elastic
and inelastic collisions.

\subsection{Neutrinos}

Neutrinos interact with the electromagnetic plasma via weak
interactions. A comprehensive treatment of this can for instance
be found in Ref.~\cite{hm95}. The collision integrals can be
written as \cite{hm95}
\begin{eqnarray}
C_{\rm coll,i} (f_1) & = & \frac{1}{2E_1} \int \frac{d^3 {\bf
p}_2}{2E_2 (2\pi)^3} \frac{d^3 {\bf p}_3}{2E_3 (2\pi)^3} \frac{d^3
{\bf p}_4}{2E_4 (2\pi)^3} \nonumber \\
&& \,\, \times (2\pi)^4 \delta^4 (p_1+p_2-p_3+p_4)
 \nonumber \\
&& \,\, \times \Lambda(f_1,f_2,f_3,f_4) S |M|_{12 \to 34,i},
\end{eqnarray}
where $S |M|_{12 \to 34,i}$ is the spin-summed and averaged matrix
element including the symmetry factor $S=1/2$ if there are
identical particles in initial or final states. The phase-space
factor is $\Lambda(f_1,f_2,f_3,f_4) = f_3 f_4 (1-f_1)(1-f_2) - f_1
f_2 (1-f_3)(1-f_4)$.

This collision integral can be reduced to 2 dimensions using the
method developed in Ref.~\cite{hm95}. However, if Pauli blocking
and interactions involving only neutrinos are neglected the
integrals can in fact be reduced to 1 dimension, as described in
Ref.~\cite{kks}. In the following we use this method. The
quantitative error resulting from this is quite small.

In addition to standard weak interactions we allow for a direct
decay of $\phi$ to neutrinos, $\phi \to \nu \bar\nu$. If $\phi$ is
non-relativistic then each neutrino is born with momentum
$m_\phi/2$ and in this case the collision integral is
\begin{equation}
C_{\phi \to \nu_i \bar\nu_i} = b_{\nu_i} \frac{2
\pi^2}{(m_\phi/2)^2} \Gamma_\phi n_\phi \delta (p_\nu - m_\phi/2),
\end{equation}
where $b_{\nu_i}$ is the branching ratio into neutrino species
$i$, and $\Gamma_\phi$ is the decay rate of the heavy particle.
For simplicity we assume equal branching ratios into all neutrino
species. Even is this is not the case the neutrino distribution
functions will be almost equilibrated by oscillations
\cite{Dolgov:2002ab}. This means that $b_{\nu_e} \simeq
b_{\nu_\mu} \simeq b_{\nu_\tau} \simeq b_\nu/3$.

Note that if one assumes that neutrinos are in kinetic equilibrium
so that they can be described by a single temperature $T_\nu$ it
is in fact possible to solve the Boltzmann equation
semi-analytically \cite{ach}. However, this is a very poor
approximation for the case when there is a direct decay mode $\phi
\to \nu \bar \nu$.

\subsection{$\phi$}

We assume the heavy particle to be completely non-relativistic. If
that is the case then the Boltzmann equation can be integrated to
give the following equation for the evolution of the energy
density
\begin{equation}
\dot \rho_\phi = -\Gamma_\phi \rho_\phi - 3 H \rho_\phi,
\label{eq:phi}
\end{equation}
i.e.\ there are no inverse decays. This is a good approximation
for all the cases covered in the present work.

We only work with masses which are low enough that there are no
hadronic decay channels open. This of course severely restricts
the possible models. However, if there is a hadronic branching
ratio then the minimum allowed reheating temperature increases
dramatically \cite{kks}, and we are investigating what the {\it
lowest} possible reheating temperature is.

\subsection{Electromagnetic plasma}

The evolution of the photon temperature can then be found from the
equation of energy conservation
\begin{equation}
\frac{d \rho_T}{d t} = -3H(\rho_T + P_T),
\end{equation}
where $\rho_T$ and $P_T$ are the total energy density and the
total pressure respectively. This equation can be rewritten as an
evolution equation for $T_\gamma$

\begin{widetext}
\begin{equation}
\frac{dT_\gamma}{dt} = -\frac{-(1-b_\nu) \rho_\phi \Gamma_\phi + 4
H \rho_\gamma + 3 H (\rho_e + P_e) + 4 H \rho_\nu + d\rho_\nu/dt}
{\partial \rho_\gamma/\partial T_\gamma + \partial \rho_e/\partial
T_\gamma} \label{eq:em}
\end{equation}
\end{widetext}

\subsection{Scale factor}

Finally we solve the Friedmann equation to find the scale factor
as a function of time
\begin{equation}
H = \frac{\dot a}{a} = \sqrt{\frac{8 \pi G \rho_T}{3}}
\label{eq:fried}
\end{equation}

Altogether we solve Eq.~(\ref{eq:fried}) together with
Eq.~(\ref{eq:boltz}) for each neutrino species, Eq.~(\ref{eq:phi})
for $\phi$, and Eq.~(\ref{eq:em}) for the photon temperature, to
obtain $a(t)$, $T_\gamma(t)$, $\rho_\phi(t)$, and $f_{\nu_i}(t)$.

\subsection{Initial conditions}

Following convention we define the reheating temperature of the
universe to be when
\begin{equation}
\Gamma_\phi = 3 H(T_{\rm RH})
\end{equation}
To a reasonable approximation the universe is radiation dominated
at this point so that
\begin{equation}
H = \left(\frac{g_* \pi^2}{90}\right)^{1/2} \frac{T_{\rm
RH}^2}{M_{\rm Pl}},
\end{equation}
where $M_{\rm Pl} = 2.4 \times 10^{18}$ GeV is the reduced Planck
mass and $g_*$ is the number of degrees of freedom.

This means that there is a one to one correspondence between
$\Gamma_\phi$ and $T_{\rm RH}$,
\begin{equation}
T_{\rm RH, MeV} \simeq 0.7 \, \Gamma_{{\rm s}^{-1}}^{1/2},
\end{equation}
where $g_* = 10.75$ has been used. Note that the constant of
proportionality is somewhat arbitrary (although it should always
be of order 1), and just gives a rough idea about the thermal
temperature when the universe enters the standard radiation
dominated phase.

As long as the initial time is set so that $t_i \ll t (T_{\rm
RH})$ {\it and} $T_{\rm max} \gtrsim T_{{\rm D},\nu}$, where
$T_{\rm max}$ is the maximum temperature reached by the plasma
after time $t_i$ and $T_{{\rm D},\nu}$ is the neutrino decoupling
temperature then the final outcome is independent of initial
conditions. The universe starts out being strongly matter
dominated and the final neutrino energy density, as well as the
light element abundances depend only on $\Gamma_\phi$, $m_\phi$,
and $b_\nu$. The initial time is found from the Friedmann equation
by assuming complete domination of $\phi$ so that $t_i =
\frac{2}{3} [8 \pi G \rho_{\phi,i}/3]^{-1/2}$.

\subsection{Nucleosynthesis}

One of the main observables from the epoch around neutrino
decoupling is the abundance of light elements, mainly helium and
deuterium. In order to calculate these abundances we have modified
the Kawano nucleosynthesis code \cite{kawano}. First is has been
modified to incorporate the modified temperature evolution, and
second the subroutines used to calculate weak interaction rates
for $n \leftrightarrow p$ have been modified to incorporate the
full numerical electron neutrino distribution coming from the
solution of the coupled Boltzmann equations.

This allows us to calculate the abundance of $^4$He and D for the
various models.

%%%%%%%%%%%%%%%%%%% section %%%%%%%%%%%%%%%%%%%%%%%%%%%%%%%%%%%%%%

\section{Numerical results}

We have solved the set of coupled Boltzmann equations for all
species for the free parameters, $m_\phi$, $\Gamma_\phi$, and
$b_\nu$.

The main output from this is the relativistic energy density in
neutrinos, parameterized in units of the energy density of a
standard model neutrino, $\rho_{\nu_0}$,
\begin{equation}
N_\nu = \frac{ \rho_{\nu_e} + \rho_{\nu_\mu} +
\rho_{\nu_\tau}}{\rho_{\nu_0}}
\end{equation}

\subsection{$b_\nu=0$}

\begin{figure}[t]
\vspace*{-0.0cm}
\begin{center}
\epsfysize=7truecm\epsfbox{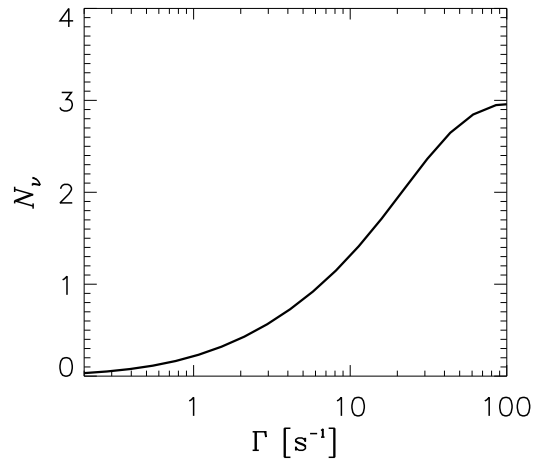}
%\vspace{0truecm}
\end{center}
\caption{The effective number of neutrino species as a function of
$\Gamma_\phi$ when there is no direct decay into neutrinos,
$b_\nu=0$.} \label{fig:nnu1}
\end{figure}

If $b_\nu=0$ then the equations become independent of $m_\phi$ and
this case has already been covered in Ref.~\cite{kks}. We present
this as our first case in order to compare results with those of
\cite{kks}. Fig.~\ref{fig:nnu1} shows the effective number of
neutrino species, $N_\nu$, after complete decay of $\phi$. This
figure is identical to Fig.~4 in Ref.~\cite{kks}.

\begin{figure}[t]
\vspace*{-0.0cm}
\begin{center}
\epsfysize=11truecm\epsfbox{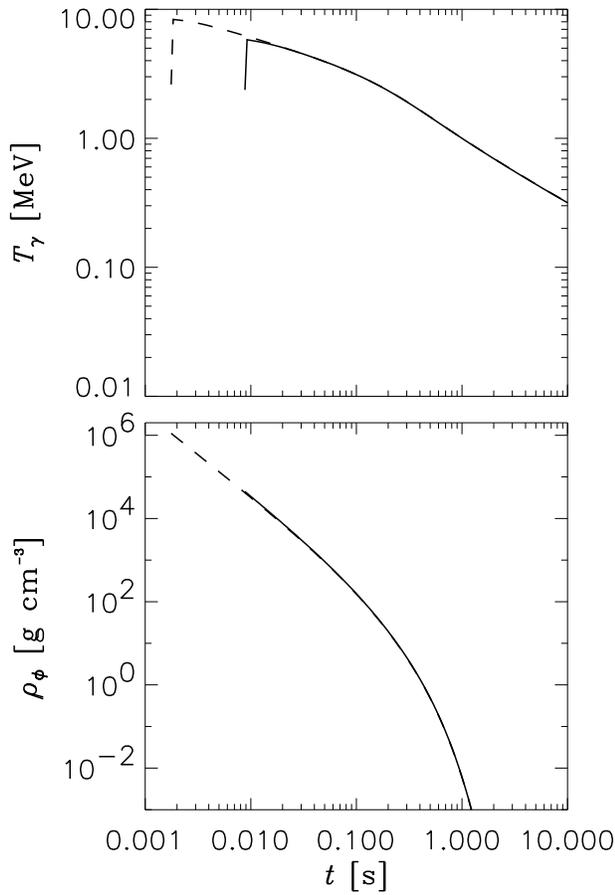} \vspace{1truecm}
\end{center}
\caption{$T_\gamma$ and $\rho_\phi$ as functions of time for
$\Gamma_\phi=6.4 \,\, {\rm s}^{-1}$, $b_\nu=0$ and two different
initial times. The full line is for $t_i=8.8 \times 10^{-3}$ s,
whereas the dashed is for $t_i=1.8 \times 10^{-3}$ s.}
\label{fig:ic}
\end{figure}

We also test whether our results are independent of initial
conditions. In Fig.~\ref{fig:ic} we show $T_\gamma(t)$ and
$\rho_\phi(t)$ for $\Gamma_\phi=6.4 \,\, {\rm s}^{-1}$ for two
different initial times, $t_i=1.8 \times 10^{-3}$ s and $t_i=8.8
\times 10^{-3}$ s. In both cases we assume an initial photon
temperature of 2.3 MeV (we could equally well have chosen an
initial temperature of 0). While the maximum temperature reached
is clearly dependent on $t_i$, $T_\gamma$ and $\rho_\phi$ quickly
become indistinguishable, and as long as the temperature where
this happens is greater than the neutrino decoupling temperature
all final results are independent of $t_i$. Furthermore, as
expected \cite{kks}, the photon temperature scales as $T_\gamma
\propto t^{-1/4}$ during the matter dominated period and shifts to
the usual $T_\gamma \propto t^{-1/2}$ once the universe becomes
radiation dominated (except for a small deviation due to heating
by $e^+e^-$ annihilation).

\subsection{$b_\nu \neq 0$}

\begin{figure*}[t]
\vspace*{-0.0cm}
\begin{center}
\epsfysize=12truecm\epsfbox{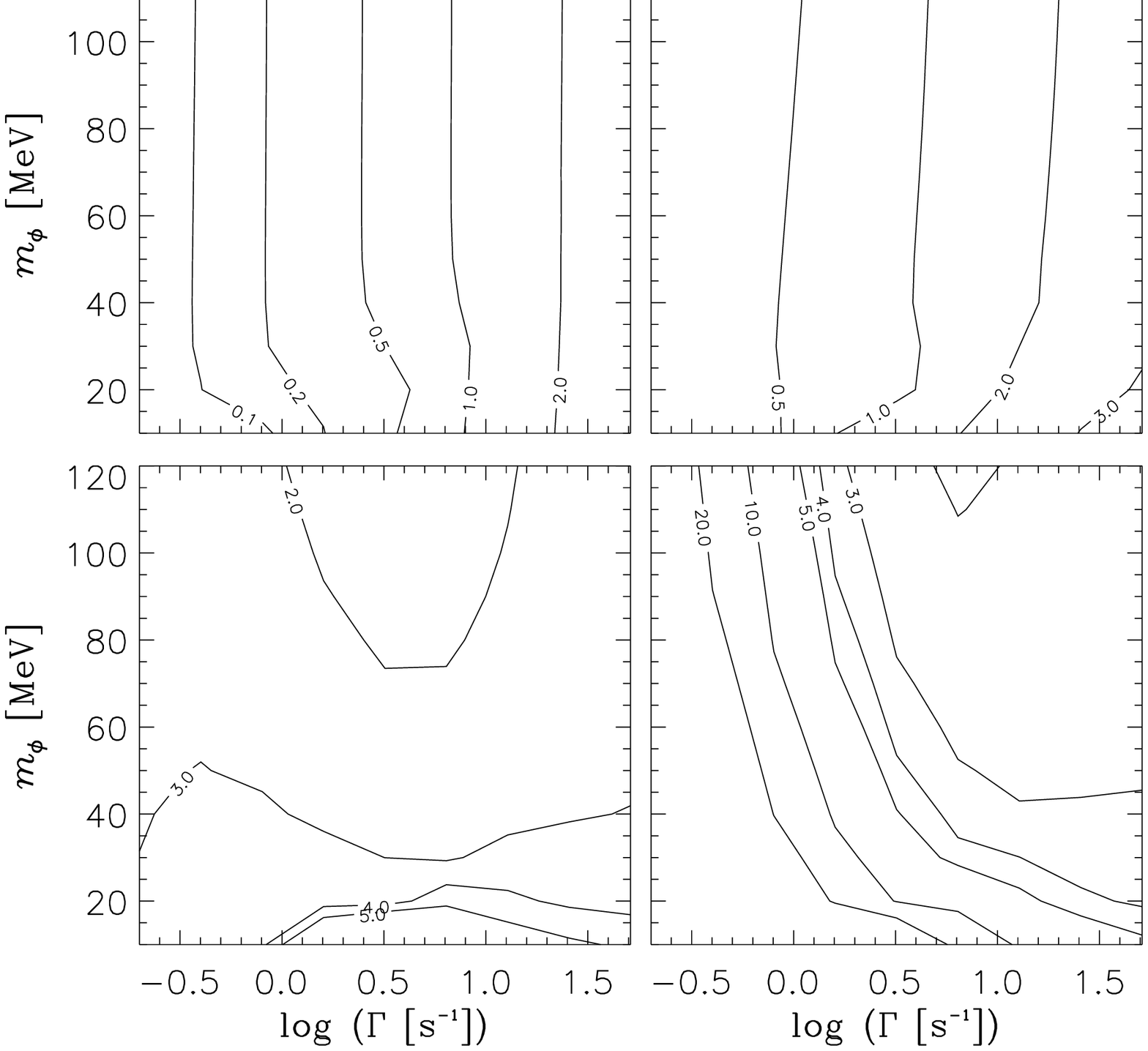}

\vspace{0.9truecm}
\end{center}
\caption{Contour plot of $N_\nu$ for different $m_\phi$ and
$\Gamma_\phi$. The top left plot is for $b_\nu=0.1$, the top right
for $b_\nu=0.5$, the bottom left for $b_\nu=0.9$, and the bottom
right for $b_\nu=1.0$.} \label{fig:nnu2}
\end{figure*}

Next we cover the case when $b_\nu \neq 0$. This is much more
complicated  to solve numerically because of the presence of the
delta function $\delta(p_\nu - m_\phi/2)$ and the fact that the
solution now depends on both $b_\nu$ and $m_\phi$. In
Fig.~\ref{fig:nnu2} we show $N_\nu$ for different values of
$\Gamma_\phi$ and $b_\nu$.

From this figure is clear that when $b_\nu$ is small the effective
number of neutrino species becomes independent of $m_\phi$ and
increasing with $\Gamma_\phi$, with $N_\nu \to 3$ for $\Gamma_\phi
\to \infty$.

For the opposite case when $b_\nu = 1$ (only decay to neutrinos)
the situation is the opposite. When $\Gamma \to \infty$ the
limiting value is again $N_\nu =3$. This corresponds to the case
when $\phi$ decays into neutrinos, but the effective neutrino
temperature after complete $\phi$ decay is higher than $T_{\rm
D}$.

When $\Gamma \to 0$ the effective number of neutrino species goes
to infinity. This corresponds to the case when $\phi$ decays so
slowly that the produced neutrinos never equilibrate with the
electromagnetic plasma, leaving only neutrinos.

However, there is a large intermediate region where $N_\nu < 3$,
even for $b_\nu = 1$. The reason for this unexpected feature can
be explained as follows: When high energy neutrinos $(E \gg T)$
are produced by direct $\phi$ decay they have a very high
annihilation cross section to $e^+ e^-$, because the cross section
goes as $E^2$. However, the produced electrons and positrons are
immediately converted into a sea of low energy $e^+$, $e^-$, and
$\gamma$ because of electromagnetic interactions. This means that
the production rate of neutrinos is much lower. In the case where
the reheating temperature is very high this does not matter
because the universe still has time to thermalize completely after
$\phi$ decay. However, if $T_{\rm RH} \sim T_{\rm D}$, this is not
possible and the result is that $N_\nu < 3$ because of the very
efficient conversion of neutrinos into $e^+ e^-$. Notice also that
this effect becomes less pronounced when $m_\phi$ decreases
because neutrinos are born with energies closer to $3T$, and the
mismatch between forward and backward rates becomes smaller.

\begin{figure}[t]
\vspace*{-0.0cm}
\begin{center}
\epsfysize=7truecm\epsfbox{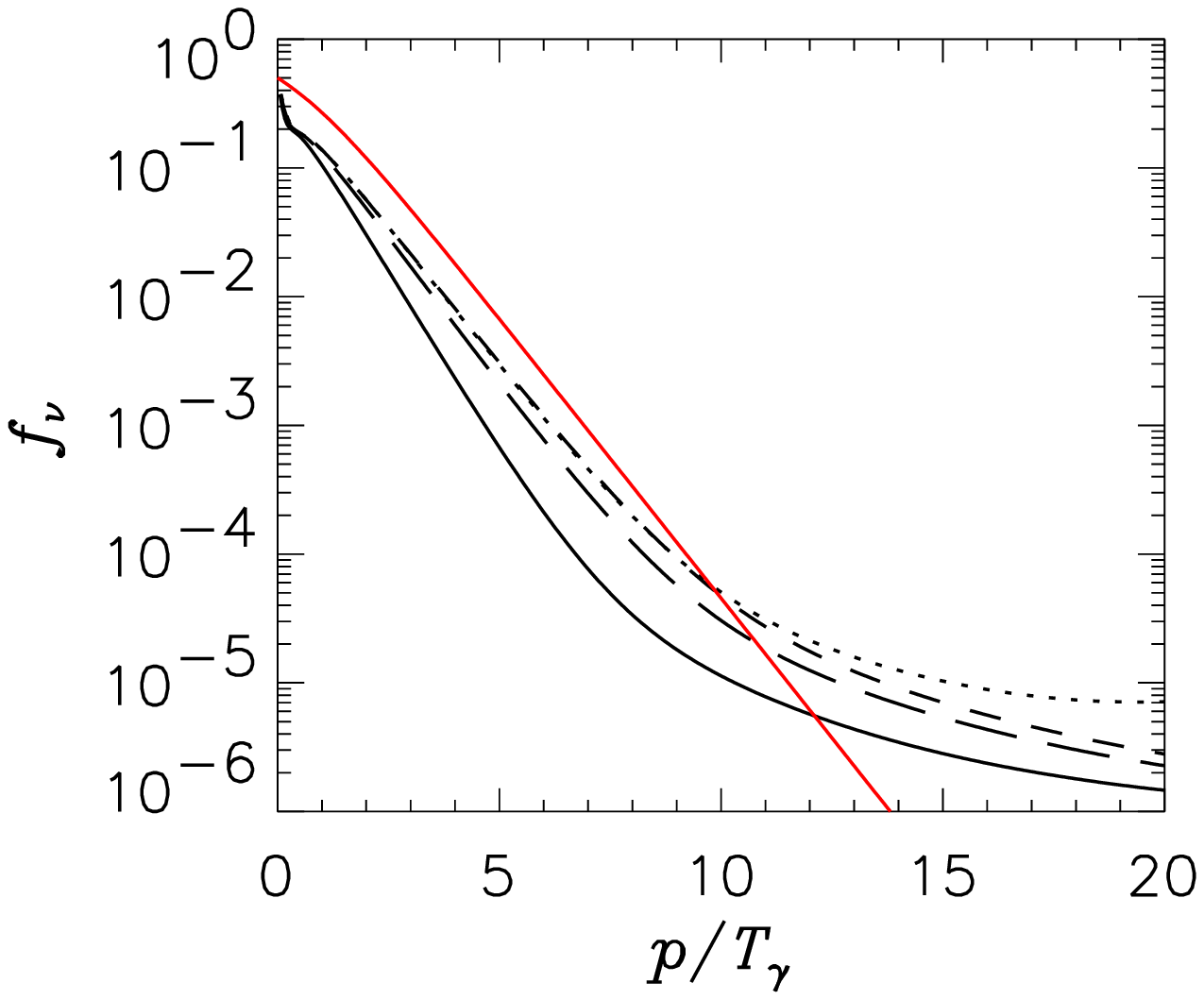}
%\vspace{0truecm}
\end{center}
\caption{The distribution function for $\nu_e$ for different
values of $T_\gamma$ when $\Gamma=6.4 \,\, {\rm s}^{-1}$,
$b_\nu=1$, and $m_\phi=120$ MeV. The dotted line is for
$T_\gamma=2.18$ MeV, the dashed for $T_\gamma=0.42$ MeV, the
long-dashed for $T_\gamma=0.19$ MeV, and the full line for
$T_\gamma=0.01$ MeV. The full grey (red) line is an equilibrium
distribution with $T_\nu=T_\gamma$.} \label{fig:dist1}
\end{figure}

\begin{figure}[t]
\vspace*{-0.0cm}
\begin{center}
\epsfysize=7truecm\epsfbox{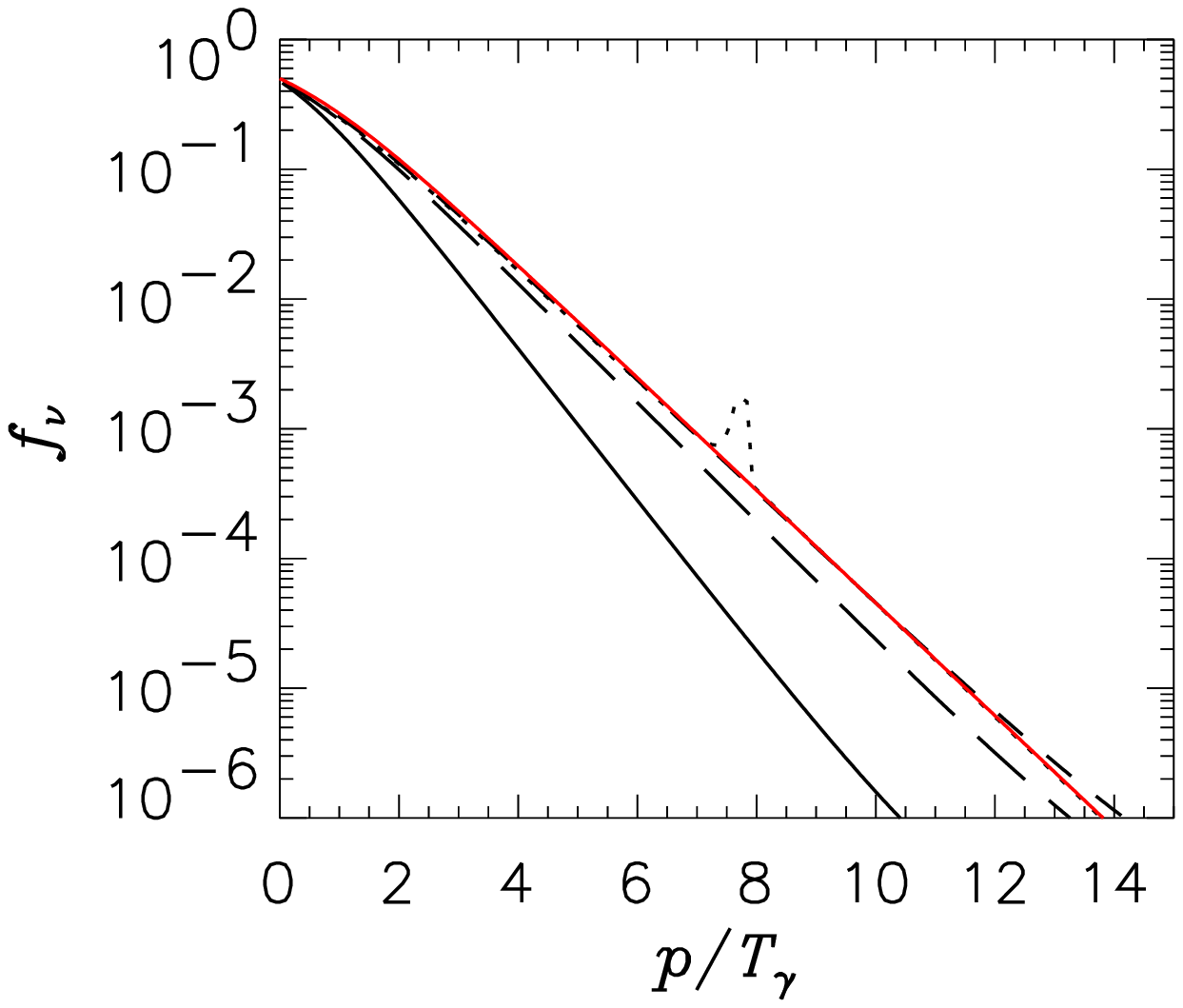}
%\vspace{0truecm}
\end{center}

\caption{The distribution function for $\nu_e$ for different
values of $T_\gamma$ when $\Gamma=50 \,\, {\rm s}^{-1}$,
$b_\nu=1$, and $m_\phi=120$ MeV. The dotted line is for
$T_\gamma=7.7$ MeV, the dashed for $T_\gamma=0.93$ MeV, the
long-dashed for $T_\gamma=0.23$ MeV, and the full line for
$T_\gamma=0.01$ MeV. The full grey (red) line is an equilibrium
distribution with $T_\nu=T_\gamma$.} \label{fig:dist2}
\end{figure}

In Figs.~\ref{fig:dist1} and \ref{fig:dist2} this effect can be
seen directly on the distribution functions. In
Fig.~\ref{fig:dist1}, which shows $b_\nu=1$, $\Gamma_\phi = 6.4
\,\, {\rm s}^{-1}$, and $m_\phi = 120$ MeV, it can be seen that
the distribution function is higher than thermal at high energies
because of $\phi$ decay. However, there are fewer low energy
neutrinos because of the inefficient production via $e^+ e^-$
annihilation.

Conversely, in Fig.~\ref{fig:dist2}, which shows $b_\nu=1$,
$\Gamma_\phi = 50 \,\, {\rm s}^{-1}$, and $m_\phi = 120$ MeV, it
can be seen that the decay rate is high enough that neutrinos
equilibrate with the electromagnetic plasma, except for a small
deviation around $p_\nu = m_\phi/2$. This subsequently leads to
$N_\nu \simeq 3$ after complete $\phi$ decay.

%%%%%%%%%%%%%%%% section %%%%%%%%%%%%%%%%%%%%%%%%%%%%%%%%%%%%%%%%%%

\section{Comparison with data}

In order to constrain the parameters $b_\nu$, $\Gamma_\phi$, and
$m_\phi$ we compare the predicted values of $N_\nu$, $^4$He, and D
with the observationally determined values. In addition to the
parameters directly related to $\phi$ the nucleosynthesis outcome
depends crucially on the baryon density, $\eta = n_B/n_\gamma$.

Taken at face value the recent CMB data from the WMAP satellite
constrain $\eta$ tightly. However, it has been shown that there is
a significant correlation between $\eta$ and $N_\nu$ in the CMB
data. This means that it is not possible to take CMB constraint on
$\eta$ directly and apply it to the nucleosynthesis calculations.
Rather a full CMB likelihood analysis for $N_\nu$ and $\eta$ must
be carried out. This can then be combined with the nucleosynthesis
likelihood analysis for $b_\nu$, $\Gamma_\phi$, $m_\phi$, and
$\eta$.

First the following subsection covers the current observational
status, then the next covers the constraints on decay parameters
which can be obtained.

\subsection{Observational data}

\subsubsection{Light element abundances}

The primordial helium abundance has been derived by two
independent groups. Fields and Olive \cite{fo} find the value
\begin{equation}
Y_P = 0.238 \pm 0.002 \pm 0.005,
\end{equation}
whereas Izotov and Thuan \cite{it} find
\begin{equation}
Y_P = 0.244 \pm 0.002 \pm 0.005
\end{equation}

Because of this inconsistency we blow up the error bars on $Y_P$
and use the value
\begin{equation}
Y_P = 0.238 \pm 0.015,
\end{equation}
which encompasses the allowed regions of both observational
determinations.

The most recent determination of the primordial deuterium
abundance has yielded the value \cite{deuterium}
\begin{equation}
{\rm D/H} = (2.78 \pm 0.29) \times 10^{-5}
\end{equation}

A determination of the primordial lithium abundance has also been
performed by several groups. However, this measurement is prone to
large systematics and we refrain from using it here.

\subsubsection{Cosmic microwave background}

The CMB temperature
fluctuations are conveniently described in terms of the
spherical harmonics power spectrum
\begin{equation}
C_l \equiv \langle |a_{lm}|^2 \rangle,
\end{equation}
where
\begin{equation}
\frac{\Delta T}{T} (\theta,\phi) = \sum_{lm} a_{lm}Y_{lm}(\theta,\phi).
\end{equation}
Since Thomson scattering polarizes light there are additional power spectra
coming from the polarization anisotropies. The polarization can be
divided into a curl-free $(E)$ and a curl $(B)$ component, yielding
four independent power spectra: $C_{T,l}, C_{E,l}, C_{B,l}$ and
the temperature $E$-polarization cross-correlation $C_{TE,l}$.

The WMAP experiment have reported data  on $C_{T,l}$ and
$C_{TE,l}$, as described in Ref.~\cite{map1,map2,map3}

We have performed the likelihood analysis using the prescription
given by the WMAP collaboration which includes the correlation
between different $C_l$'s \cite{map1,map2,map3}. Foreground
contamination has already been subtracted from their published
data.

\subsubsection{Large scale structure}

The 2dF Galaxy Redshift Survey (2dFGRS) \cite{colless} has measured
the redshifts
of more than 230 000 galaxies with a median redshift of
$z_{\rm m} \approx 0.11$.
An initial estimate of the convolved, redshift-space power spectrum of the
2dFGRS has been determined \cite{percival} for a sample of 160 000 redshifts.
On scales $0.02 < k < 0.15h\;{\rm Mpc}^{-1}$ the data are robust and the
shape of the power spectrum is not affected by redshift-space or nonlinear
effects, though the amplitude is increased by redshift-space distortions.
A potential complication is the fact that the galaxy power spectrum
may be biased with respect to the matter power spectrum, i.e. light does not
trace mass exactly at all scales.  This is often parametrised by introducing
a bias factor
\begin{equation}
b^2(k)\equiv \frac{P_{\rm g}(k)}{P_{\rm m}(k)},
\label{eq:biasdef}
\end{equation}
where $P_{\rm g}(k)$ is the power spectrum of the galaxies, and
$P_{\rm m}(k)$ is the matter power spectrum. However, we restrict
our analysis of the 2dFGRS power spectrum to scales $k < 0.15
\;{\rm h}\,{\rm Mpc}^{-1}$ where the power spectrum is well
described by linear theory. On these scales, two different
analyses have demonstrated that the 2dFGRS power spectrum is
consistent with linear, scale-independent bias \cite{lahav,verde}.
Thus, the shape of the galaxy power spectrum can be used
straightforwardly to constrain the shape of the matter power
spectrum.

The only parameters which affect CMB and structure formation are
the baryon density, $\eta$, and the relativistic energy density at
late times, parameterized by $N_\nu$ \cite{hannestad03,barger}
(see also \cite{Hannestad:2000hc,Esposito:2000sv,Kneller:2001cd,%
Hannestad:2001hn,Hansen:2001hi,Bowen:2001in,pastor,pierpaoli}). It
is therefore relatively straightforward to perform the CMB+LSS
likelihood analysis.

\subsection{Likelihood analysis}

Nucleosynthesis is affected both by the expansion rate around $T
\sim 0.1-1$ MeV, and by the electron neutrino distribution
function. The reason is that electron neutrino enter directly in
the weak reactions which interconvert protons and neutrons.

The specific neutrino distributions are therefore found as
functions of temperature and used in a modified version of the
Kawano BBN code \cite{kawano}. This is then used to calculate
primordial abundances of deuterium and helium.

For calculating the theoretical CMB and matter power spectra we
use the publicly available CMBFAST package \cite{CMBFAST}. As the
set of cosmological parameters we choose $\Omega_m$, the matter
density, $\Omega_b$, the baryon density, $H_0$, the Hubble
parameter, $\tau$, the optical depth to reionization, $Q$, the
normalization of the CMB power spectrum, $b$, the bias parameter,
and the effective number of neutrino species $N_\nu$, found from
the solution of the Boltzmann equations. We assume neutrinos to be
almost massless. We restrict the analysis to geometrically flat
models $\Omega_m + \Omega_\Lambda = 1$.

For each individual model we calculate $\chi^2$ in the following
way: Given a theoretical CMB spectrum the $\chi^2$ of the WMAP
data is calculated using the method described in Ref.~\cite{map3}.
With regards to the 2dF data we use the data points and window
functions from Ref.~\cite{Tegmark:2001jh}
(http://www.hep.upenn.edu/$\sim$max/2df.html). 68\% and 95\%
confidence levels from the data are calculated from $\Delta \chi^2
= 2.31$ and 6.17 respectively.

\subsubsection{$b_\nu=0$}

\begin{figure}[t]
\vspace*{-0.0cm}
\begin{center}
\epsfysize=11truecm\epsfbox{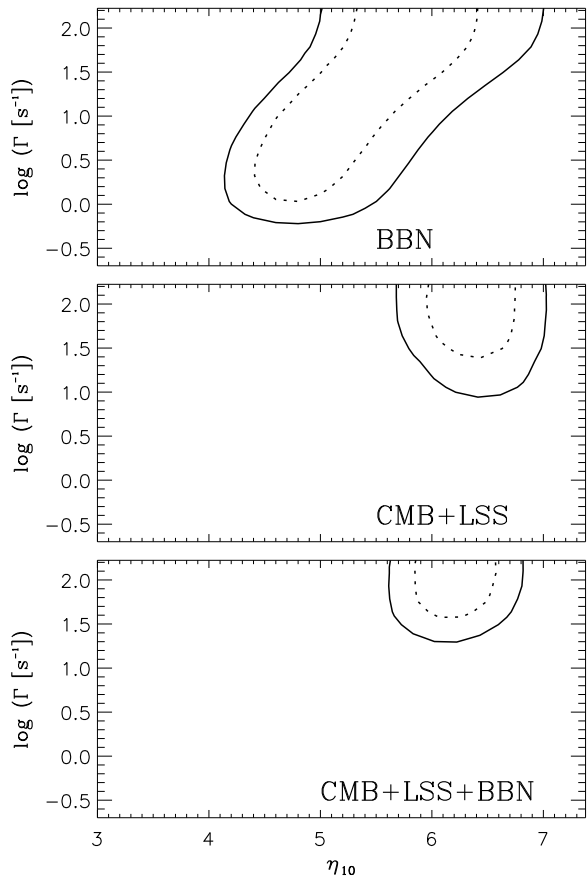}

\vspace{2.5truecm}
\end{center}
\caption{68\% and 95\% confidence exclusion plot of the parameters
$\eta_{10} \equiv 10^{10} \times \eta$ and $\Gamma_\phi$ for the
case when $b_\nu=0$.} \label{fig:excl1}
\end{figure}

In Fig.~\ref{fig:excl1} we show 68\% and 95\% exclusion limits for
$\eta$ and $\Gamma_\phi$ from BBN, CMB, and LSS. The top panel for
BBN only is very similar to Fig.~8 in KKS, except that we use
slightly different bounds on light element abundances. From BBN
alone the 95\% bound on $T_{\rm RH}$ is roughly 0.6 MeV. However
this bound is achieved for relatively low $\eta$, whereas CMB+LSS
strongly prefer a high value of $\eta$. Therefore combining the
BBN and CMB+LSS constraints removes the low $T_{\rm RH}$ region
and increases the lower bound to 3.9 MeV.

\begin{figure*}[t]
\vspace*{-0.0cm}
\begin{center}
\epsfysize=12truecm\epsfbox{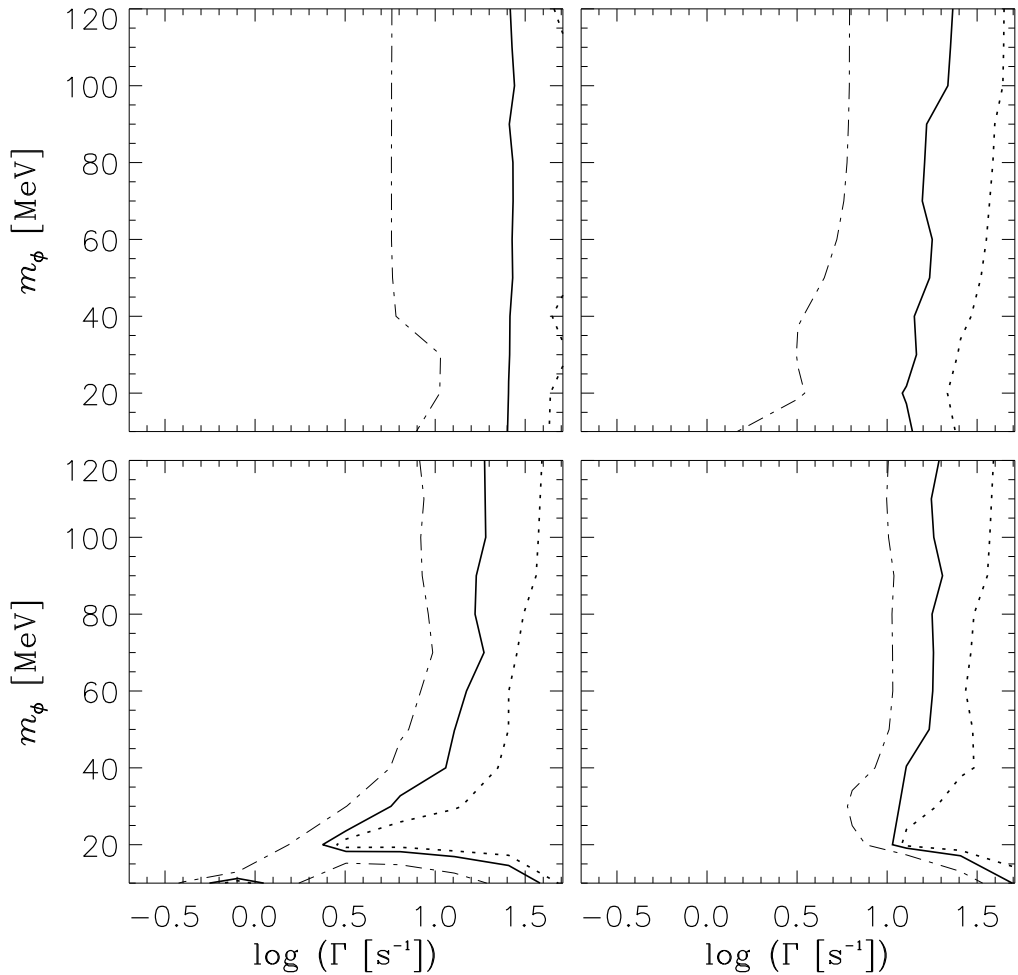}

\vspace{0.9truecm}
\end{center}
\caption{68\%, 95\%, and 99.99\% confidence exclusion plot of the
parameters $\Gamma_\phi$ and $m_\phi$ using all available data
(CMB+LSS+BBN). The top left plot is for $b_\nu=0.1$, the top right
for $b_\nu=0.5$, the bottom left for $b_\nu=0.9$, and the bottom
right for $b_\nu=1.0$.} \label{fig:excl2}
\end{figure*}

\subsubsection{$b_\nu \neq 0$}

Apart from the fact that $N_\nu$ depends on $b_\nu$ there is a
second effect which is just a important. When $b_\nu \neq 0$ there
are more high energy neutrinos. Around weak freeze-out there are
many more protons than neutrons. When $E_\nu \gg m_n-m_p$ the weak
absorption cross section is equal on protons and neutrons. This
means that additional neutrinos at high energies will have the net
effect of converting protons into neutrons, so that in the end
more helium is produced. Note that this is the opposite effect of
just increasing the weak interaction rates, in which case {\it
less} helium would be produced. The phenomenon is quite similar to
what happens if $\phi$ has a hadronic decay channel. In that case
pions and kaons will be produced, which subsequently convert
protons to neutrons and lead to overproduction of helium.

In Fig.~\ref{fig:excl2} we show 68\%, 95\%, and 99.99\% confidence
exclusion plots for $\Gamma_\phi$ and $m_\phi$, marginalized over
$\eta$.

Both when $b_\nu$ is small and when $b_\nu=1$ the bound on $T_{\rm
RH}$ becomes independent of $m_\phi$. In both cases the 95\% bound
is $T_{\rm RH} \gtrsim 4$ MeV.

However, there is an intermediate regime for $b_\nu$ which allows
for much lower values of $T_{\rm RH}$. The reason for this can be
seen directly from Fig.~\ref{fig:nnu2}, i.e. there is an
intermediate range where $N_\nu$ can be kept close to 3, even for
low $\Gamma_\phi$. However, for large masses (which is of course
by far the most likely) there is no allowed region. The reason is
the one given in the previous section: More high energy neutrinos
will produce more helium, and this in turn will conflict with
observations.

The final outcome is that for almost all values of $m_\phi$ and
$b_\nu$ there is a robust lower bound on $T_{\rm RH}$ which is
around 4 MeV. However there is a small region where $b_\nu \sim
0.9$, $m_\phi \lesssim 40$ MeV where a reheating temperature as
low as roughly 1 MeV is allowed.

%%%%%%%%%%%%%%%%%%% section %%%%%%%%%%%%%%%%%%%%%%%%%%%%%%%%%%%%%

\section{Other constraints}

If $\phi$ is a scalar then the decay rate to neutrinos is normally
suppressed by a factor $m_\nu^2$ because of the necessary helicity
flip. Therefore the simplest assumption is that $\phi$ has no
branching into neutrinos. If for instance the heavy particle is a
pseudo-scalar like the axion, then there is an upper bound on the
coupling to photons (\cite{pdg,raffelt}), $g_{\phi \gamma} \leq
0.6 \times 10^{-10} \,\, {\rm GeV}^{-1}$ for $m_\phi \lesssim 30$
keV. For higher masses the bound is significantly weaker. However,
even if this bound is used together with the decay width
$\Gamma_{\phi \to 2 \gamma} = g_{\phi \gamma}^2 m_\phi^3/64\pi$
then we find that
\begin{equation}
\Gamma_{\phi \to 2 \gamma} \lesssim 50 \, m_{\phi,10 \, {\rm
MeV}}^3 \,\, {\rm s}^{-1},
\end{equation}
which is easily satisfies for the parameter space we are
considering.

On the other hand, if $\phi$ is a particle like the majoron which
couples only to neutrinos then the decay width is \cite{bv}
\begin{equation}
\Gamma_{\phi \to \nu \bar\nu} = \frac{g_{\phi \nu}^2 m_\phi}{16
\pi} \sim 3 \times 10^{19} \, g_{\phi \nu}^2 \, m_{\phi, {\rm
MeV}} \,\, {\rm s}^{-1}
\end{equation}
The bound on the dimensionless coupling constant comes from BBN as
well as supernova considerations and is of order $10^{-6} -
10^{-5}$ for majorons in the MeV mass range \cite{raffelt,valle}.
For more massive majorons the bound weakens. Again it is clear
that the decay parameters which we consider here are not excluded
by any other astrophysical or experimental data.

The final conclusion is that heavy, decaying particles such as the
ones considered here cannot be directly excluded by any current
data. Furthermore a branching ratio into neutrinos can be anywhere
from 0 to 1.

%%%%%%%%%%%%%%%%%%% section %%%%%%%%%%%%%%%%%%%%%%%%%%%%%%%%%%%%%

\section{Conclusion}

We have carefully calculated constraints on models with extremely
low reheating temperature, where a massive particle decays around
$T \sim 1$ MeV. By combining constraints on light element
abundances with constraints on $\eta$ and $N_\nu$ from CMB and
large scale structure we derived a fairly robust limit of
\begin{equation}
T_{\rm RH} \gtrsim 4 \,\,\, {\rm MeV}.
\end{equation}
This bound is a significant improvement over the previous bound of
$T_{\rm RH} \gtrsim 0.7$ MeV, calculated from BBN alone. It is
interesting that the lower bound is significantly higher than the
$n \leftrightarrow p$ conversion freeze-out temperature, $T \sim
0.8$ MeV, end even higher than the neutrino decoupling temperature
$T_{\rm D} \sim 2$ MeV. This shows that even small residual
effects can be measured with present observational data.

Models with reheating temperature in the MeV regime are in general
difficult to reconcile with such features as baryogenesis.
However, in models with large extra dimensions a low reheating
temperature is essential in order to avoid overproduction of
massive Kaluza-Klein graviton states. This means that we can use
our present bound to derive limits on the compactification scale
in such models. For the case of two extra dimensions the bound is
$M \gtrsim 2000$ TeV and for $n=3$ it is $M \gtrsim 100$ TeV. This
bound is somewhat stronger than the bound coming from
considerations of neutron star cooling and gamma ray emission.

\section*{Acknowledgments}

We acknowledge use of the publicly available CMBFAST package
written by Uros Seljak and Matias Zaldarriaga \cite{CMBFAST}, as
well as the nucleosynthesis code written by Lawrence Kawano
\cite{kawano}. I wish to thank P.~Serpico for comments.

\newcommand\AJ[3]{~Astron. J.{\bf ~#1}, #2~(#3)}
\newcommand\APJ[3]{~Astrophys. J.{\bf ~#1}, #2~ (#3)}
\newcommand\apjl[3]{~Astrophys. J. Lett. {\bf ~#1}, L#2~(#3)}
\newcommand\ass[3]{~Astrophys. Space Sci.{\bf ~#1}, #2~(#3)}
\newcommand\cqg[3]{~Class. Quant. Grav.{\bf ~#1}, #2~(#3)}
\newcommand\mnras[3]{~Mon. Not. R. Astron. Soc.{\bf ~#1}, #2~(#3)}
\newcommand\mpla[3]{~Mod. Phys. Lett. A{\bf ~#1}, #2~(#3)}
\newcommand\npb[3]{~Nucl. Phys. B{\bf ~#1}, #2~(#3)}
\newcommand\plb[3]{~Phys. Lett. B{\bf ~#1}, #2~(#3)}
\newcommand\pr[3]{~Phys. Rev.{\bf ~#1}, #2~(#3)}
\newcommand\prog[3]{~Prog. Theor. Phys.{\bf ~#1}, #2~(#3)}

\end{document}